\documentstyle[psfig,12pt,citesort]{article}

\def \ov {\over}

\def \P { \Phi} \def\ep {\epsilon}

\def \F {{\cal F}}

\def \p {\phi}
\def \ep {\epsilon}
\def \te {\tilde \epsilon}
\def \ps {\psi}

\newcounter{subequation}[equation]

\newcommand{\be}{\begin{equation}}
\newcommand{\ee}{\end{equation}}
\newcommand{\eel}[1]{\label{#1}\end{equation}}
\newcommand{\bea}{\begin{eqnarray}}
\newcommand{\eea}{\end{eqnarray}}
\newcommand{\eeal}[1]{\label{#1}\end{eqnarray}}

\makeatletter

\def\thesubequation{\theequation\@alph\c@subequation}
\def\@subeqnnum{{\rm (\thesubequation)}}
\def\slabel#1{\@bsphack\if@filesw {\let\thepage\relax
   \xdef\@gtempa{\write\@auxout{\string
      \newlabel{#1}{{\thesubequation}{\thepage}}}}}\@gtempa
   \if@nobreak \ifvmode\nobreak\fi\fi\fi\@esphack}
\def\subeqnarray{\stepcounter{equation}
\let\@currentlabel=\theequation\global\c@subequation\@ne
\global\@eqnswtrue \global\@eqcnt\z@\tabskip\@centering\let\\=\@subeqncr

$$\halign to \displaywidth\bgroup\@eqnsel\hskip\@centering
  $\displaystyle\tabskip\z@{##}$&\global\@eqcnt\@ne
  \hskip 2\arraycolsep \hfil${##}$\hfil
  &\global\@eqcnt\tw@ \hskip 2\arraycolsep
  $\displaystyle\tabskip\z@{##}$\hfil
   \tabskip\@centering&\llap{##}\tabskip\z@\cr}
\def\endsubeqnarray{\@@subeqncr\egroup
                     $$\global\@ignoretrue}
\def\@subeqncr{{\ifnum0=`}\fi\@ifstar{\global\@eqpen\@M
    \@ysubeqncr}{\global\@eqpen\interdisplaylinepenalty \@ysubeqncr}}
\def\@ysubeqncr{\@ifnextchar [{\@xsubeqncr}{\@xsubeqncr[\z@]}}
\def\@xsubeqncr[#1]{\ifnum0=`{\fi}\@@subeqncr
   \noalign{\penalty\@eqpen\vskip\jot\vskip #1\relax}}
\def\@@subeqncr{\let\@tempa\relax
    \ifcase\@eqcnt \def\@tempa{& & &}\or \def\@tempa{& &}
      \else \def\@tempa{&}\fi
     \@tempa \if@eqnsw\@subeqnnum\refstepcounter{subequation}\fi
     \global\@eqnswtrue\global\@eqcnt\z@\cr}
\let\@ssubeqncr=\@subeqncr
\@namedef{subeqnarray*}{\def\@subeqncr{\nonumber\@ssubeqncr}\subeqnarray}

\@namedef{endsubeqnarray*}{\global\advance\c@equation\m@ne%
                           \nonumber\endsubeqnarray}

\makeatletter \@addtoreset{equation}{section} \makeatother
\renewcommand{\theequation}{\thesection.\arabic{equation}}

\catcode`\@=11

\newcount\hour
\newcount\minute
\newtoks\amorpm \hour=\time\divide\hour by 60\minute
=\time{\multiply\hour by 60 \global\advance\minute by-\hour}
\edef\standardtime{{\ifnum\hour<12 \global\amorpm={am}%
        \else\global\amorpm={pm}\advance\hour by-12 \fi
        \ifnum\hour=0 \hour=12 \fi
        \number\hour:\ifnum\minute<10
        0\fi\number\minute\the\amorpm}}
\edef\militarytime{\number\hour:\ifnum\minute<10 0\fi\number\minute}

\def\draftlabel#1{{\@bsphack\if@filesw {\let\thepage\relax
   \xdef\@gtempa{\write\@auxout{\string
      \newlabel{#1}{{\@currentlabel}{\thepage}}}}}\@gtempa
   \if@nobreak \ifvmode\nobreak\fi\fi\fi\@esphack}
        \gdef\@eqnlabel{#1}}
\def\@eqnlabel{}
\def\@vacuum{}
\def\marginnote#1{}
\def\draftmarginnote#1{\marginpar{\raggedright\scriptsize\tt#1}}
\def\draft{
        \pagestyle{plain}
        \overfullrule=2pt
        \oddsidemargin -.5truein
        \def\@oddhead{\sl \phantom{\today\quad\militarytime} \hfil
        \smash{\Large\sl DRAFT} \hfil \today\quad\militarytime}
        \let\@evenhead\@oddhead
        \let\label=\draftlabel
        \let\marginnote=\draftmarginnote
        \def\ps@empty{\let\@mkboth\@gobbletwo
        \def\@oddfoot{\hfil \smash{\Large\sl DRAFT} \hfil}
        \let\@evenfoot\@oddhead}

\def\@eqnnum{(\theequation)\rlap{\kern\marginparsep\tt\@eqnlabel}%
        \global\let\@eqnlabel\@vacuum}  }

\renewcommand{\theequation}{\thesection.\arabic{equation}}
\renewcommand{\thefootnote}{\fnsymbol{footnote}}

\def\appendix#1{
  \addtocounter{section}{-3}
  \setcounter{equation}{0}
  \renewcommand{\thesection}{\Alph{section}}
  \section*{Appendix \thesection\protect\indent \parbox[t]{11.15cm}
  {#1} }
  \addcontentsline{toc}{section}{Appendix \thesection\ \ \ #1}
  }
\def \tix{\tilde{x}}

\def \ov {\over}

\def \P { \Phi} \def\ep {\epsilon}

\def \F {{\cal F}}

\def \p {\phi}
\def \ep {\epsilon}
\def \te {\tilde \epsilon}
\def \ps {\psi}

\def\O{\Omega}

\def\m{\mu}
\def\n{\nu}
\def\a{\alpha}

\def\r{\rho}

\def\l{\lambda}
\def\te{\theta}

\def\p{\phi}
\def\L{\Lambda}

\def\ep{\epsilon}

\def\be{\begin{equation}}
\def\ee{\end{equation}}
\def \P {\Phi}

\def \k {\kappa}
\def \l {\lambda}
\def \L {{L}}

\def \m {\mu}
\def \n {\nu}
\def \W{{\cal W}}

\def \de{{
{ 1 \ov 9}} }
\def \si{{
 { 1 \ov 6}} }

\def\l{\lambda}
\def\L{\Lambda}
\def\te{\theta}

\date{}
\begin{document}

\begin{titlepage}

\hfill hep-th/0202186

\hfill MCTP-02-11\\

\begin{center}

{\Large \bf On Penrose Limits and Gauge Theories}

\vskip .7 cm

\vskip 1 cm

{\large   Leopoldo A. Pando Zayas${}^{1,2}$ and Jacob Sonnenschein${}^{2,3}$}\\

\end{center}

\vskip .4cm \centerline{\it ${}^1$ Michigan Center for Theoretical
Physics}
\centerline{ \it Randall Laboratory of Physics, The University of
Michigan}
\centerline{\it Ann Arbor, MI 48109-1120}

\vskip .4cm \centerline{\it ${}^2$ School of Natural Sciences}
\centerline{ \it Institute for Advanced Study}
\centerline{\it Princeton, NJ 08540} 

\vskip .4cm \centerline{\it ${}^3$ School of Physics and Astronomy}
\centerline{ \it Beverly and Raymond Sackler Faculty of Exact Sciences}
\centerline{ \it Tel Aviv University, Ramat Aviv, 69978, Israel}

\vskip 1 cm

\vskip 1.5 cm

\begin{abstract}
We discuss various Penrose  limits of conformal and nonconformal 
backgrounds. In $AdS_5\times T^{1,1}$, for a
particular choice of the angular coordinate in $T^{1,1}$ the 
resulting Penrose
limit coincides with the similar limit for $AdS_5\times S^5$. 
In this case an  identification of  a subset of field theory operators 
with the string zero mode creation operators is possible. For another 
limit we obtain a light-cone string action that resembles a particle in a 
magnetic field. 
We also consider three different types of backgrounds that are dual to 
nonconformal field theories: The Schwarzschild black hole 
in $AdS_5$, D3-branes
on the small resolution of the conifold and the 
Klebanov-Tseytlin background. 
We find that in all  three cases the introduction of 
nonconformality renders a theory that is no longer  exactly solvable
and that the form of the deformation is universal. The corresponding 
world sheet theory in the light-cone gauge has a $\tau=x^+$ dependent
mass term.   
\end{abstract}

\end{titlepage}

\setcounter{page}{1} \renewcommand{\thefootnote}{\arabic{footnote}}
\setcounter{footnote}{0}

\def \N{{\cal N}} \def \ov {\over}

\section{ Introduction}
Recently the idea of Penrose that ``Any space-time has a plane wave a
limit''\cite{penrose} has undergone  a period of  a ``
Renaissance''. In Penrose's own words, this limit is the adaptation to 
pseudo Riemannian manifolds of the well-understood procedure  of taking tangent
space limit. The main 
difference being that whether $T_p$ in essentially flat, the Penrose limit 
applied to a null geodesic results in curved space - a plane wave. 
The idea has been extended, generalized and applied to
the study of string theories on the corresponding backgrounds.   In
particular  several papers  have been devoted to a detailed
analysis of the application of the limit  to supergravity backgrounds
\cite{gueven,tratado,figueroa}, and to explicitly solving the
superstring theory defined on the ppw
background\cite{metsaev} as well as understanding the relation between the  
string spectrum and massless supergravity modes \cite{metse}. 
A similar limit was explored also in \cite{old}.
In \cite{bmn} the string spectrum
of a pp-wave background was shown to arise from the large N limit
( large $g^2_{YM}N$, fixed $g^2_{YM}N/J^2$) of ${\cal N}=4$ SYM theory
in 4d. This  was accomplished by summing a subset of planar
diagrams.  The work of Berenstein, Maldacena and Nastase is a very interesting
extension of the original AdS/CFT correspondence \cite{ads1} in that it 
considers massive string modes.

The goal of this paper is two folded: to apply the
Penrose limit to other conformal 
 and non conformal backgrounds and
to   
 examine the question of whether the procedure introduced in
\cite{bmn} can be used to identify
 a string spectrum in certain
classes of operators of gauge theories which have ${\cal N}=1$
supersymmetry and are non-conformal theories. We construct
various limits of the  $AdS_5\times T^{1,1}$ background.  For one 
particular choice of a geodesic line  the resulting Penrose-G\"uven limit
coincides with the similar limit for $AdS_5\times S^5$.
 Another
limit associated with a different geodesic yields a string theory
resembling a particle in a  
 magnetic field.  We then analyze the
limits of backgrounds that are not dual to  conformal field
theories: The Schwarzschild black hole in $AdS_5$, D3-branes 
 on the
small resolution of the conifold and the Klebanov-Tseytlin background.

We find that in all  these cases there exist a limit which results
in a string theory described
 by a worldsheet action that includes a
mass term that depends on the worldsheet time.
 We make the first
steps toward the identification of the field theory operators that
correspond to the low lying string states. 

While this
paper was being prepared for publication  two manuscripts \cite{ikm} and
\cite{go} that discuss similar questions appeared on the net. These
papers overlap with our discussions of the limits in the 
 $ T^{1,1}$
and $T^{p,q}$ models as well as the identification of the string
spectrum in the field theory picture. 

The paper is organized as
follows.  In section 2  various Penrose limits 
 of conformal
backgrounds are taken. In particular in the context of  $AdS_5
\times T^{1,1}$  we identify a limit which reproduce the 
exact geometry found in \cite{bmn} for the $AdS_5 \times S^5$ case and
describe another limit which leads to  a string background resembling
a particle in a magnetic field. We then discuss the general $T^{p,q}$
background.
 Section 2 is devoted to the analysis of several non
conformal cases. These include   the AdS black hole,  the small
resolution of the conifold \cite{pt} and the Klebanov Tseytlin model
\cite{kt}.  We then comment on the general structure of the Penrose
limit of the non-conformal backgrounds. In the appendix we discuss 
the local  nature of the limit we take in the context of 
conformal backgrounds.

\section{Penrose limits in conformal backgrounds}
The analysis of \cite{bmn} is completely symmetric with 
respect to the choice of the $U(1)$ coordinate inside $S^5$ or equivalently 
the $U(1)$ subgroup of $SU(4)$ R-symmetry. In the case 
of $AdS_5 \times T^{1,1}$
there is a clear difference between the three possible $U(1)$ coordinates; 
only one correspond to the $U(1)$  R-symmetry. To clarify this difference we 
next study various  possibilities. 

\subsection{PP wave limit on $AdS_5 \times T^{1,1}$}
\label{n4}
We consider various Penrose limits in the geometry of $AdS_5 \times T^{1,1}$
\begin{eqnarray}
\label{t11}
R^{-2}ds^2&=&-dt^2 \cosh^2 \r + d\r^2 + \sinh^2 \r d \O_3^2  \nonumber \\
&+&{1\over 9}\bigg[d\psi + \cos \te_1 d\p_1+ \cos \te_2 d\p_2\bigg]^2
+ {1\over 6}\bigg[d\te_1^2 + \sin^2\te_2 d\p_1^2+
d\te_1^2 +  \sin^2\te_2 d\p_1^2\bigg].
\end{eqnarray}
In analogy with \cite{bmn} we concentrate on the motion of a particle that moves
along a direction given  by $\psi +\p_1 + \p_2$ in a geodesic defined in
a small neighborhood of  
$\r=0$ and $\te_i=0$. Technically this amounts to changing the metric to new 
coordinates  
\begin{eqnarray}
\tix^+&=& {1\over 2}\bigg[t+{1\over 3}(\psi + \p_1 + \p_2)\bigg], \nonumber \\
\tix^-&=& {1\over 2}\bigg[t-{1\over 3}(\psi + \p_1 + \p_2)\bigg], \nonumber \\
\Phi_1&=&\p_1- {1\over 2}\bigg[t+{1\over 3}(\psi + \p_1 + \p_2)\bigg], \nonumber \\
\Phi_2&=&\p_2- {1\over 2}\bigg[t+{1\over 3}(\psi + \p_1 + \p_2)\bigg],
\end{eqnarray}
and subsequently we take the $R\to \infty$ limit with 
\be
x^+=\tix^+, \quad x^-=R^2 \tix^-, \quad \r={r\over R}, 
\quad \te_i = \sqrt{6} {r_i\over R}.
\ee
In this limit the metric becomes 
\begin{eqnarray}
ds^2&=&-4dx^+ dx^- - \m^2 (r^2 + r_1^2 + r_2^2)( dx^+)^2  \nonumber \\
&+& dr^2+r^2 d\O_3^2+ dr_1^2+r_1^2d\P_1^2 +dr_2^2 +r_2^2 d\P_2^2,
\end{eqnarray}
where we have introduced the mass parameter $\m$ as a rescaling 
$x^+\to \m x^+$ and $x^-\to x^-/\m$. Since each pair $(r_i, \P_i)$
parametrizes an 
$R^2$ we end up with a result exactly matching that of \cite{bmn},
that is, a background of the form
\begin{eqnarray}
\label{ppwave}
ds^2&=&-4dx^+ dx^- -\m^2 \vec{z}\,{}^2(dx^+)^2 + d\vec{z}\,{}^2, \nonumber \\
F_{+1234}&=&F_{+5678}\sim  \m.
\end{eqnarray}
This background has been studied as an exactly solvable string theory
in Ramond-Ramond background \cite{metsaev,metse}.

\subsection{Other limits of $AdS_5 \times T^{1,1}$, a  magnetic case}
\label{mag}
The combination of variables that we took in the previous subsection was rather 
particular and was dictated by field theory considerations which we will discuss
 in more detail in section \ref{field}. It is also natural, from the geometrical point
of view to consider other limits. In particular it is natural to consider a limit 
of particles moving along the $\psi$ or $\phi_i$ directions. We will see that 
they look rather different. Introducing in (\ref{t11}) the following 
change of variable 
\be
\tix^+= {1\over 2}(t+{1\over 3}\psi), \quad  
\tix^-= {1\over 2}(t-{1\over 3}\psi)
\ee
and further taking the limit
\begin{eqnarray}
x^+&=&\tix^+, \quad x^-=R^2 \tix^-, \quad \r={r\over R}, \nonumber \\
\te_i &=& {\pi\over 2}+\sqrt{6} { y_i\over R}, \quad  \p_i =
\sqrt{6}{x_i\over R}.
\end{eqnarray}
we obtain the following metric resembling the motion of a particle in a magnetic 
field. 
\begin{eqnarray}
ds^2&=&-4dx^+ (dx^- +\m y_1 dx_1 +\m y_2 dx_2)-\m^2 \vec{r}\,{}^2(dx^+)^2
+ d\vec{r}\,{}^2+ d\vec{y}{}^2 + d\vec{x}{}^2, \nonumber \\
F_5&=&\F_5+ *\F_5, \quad \F_5\sim \m \, dx^+\wedge dy^1\wedge dx^1
\wedge dy^2\wedge dx^2,
\end{eqnarray}
where $\vec{y}=(y_1,y_2)$ and $\vec{x}=(x_1,x_2)$. This background can be 
transformed into (\ref{ppwave}) by means of an appropriate coordinate 
transformation. However, from the field theory point of view (see discussion 
in subsection \ref{magnetic}) these two limits seem to be distinct. 
 
The gauge fixed light-cone bosonic string action in this case is
\be
S= {1\over 2\pi\alpha'}\int d^2 \sigma [
{1\over 2}\partial_\alpha \vec{r} \partial^\alpha \vec{r} 
+{1\over 2}\partial_\alpha \vec{y} \partial^\alpha \vec{y}
+{1\over 2}\partial_\alpha \vec{x} \partial^\alpha \vec{x}  
-{1\over 2}\mu^2\vec{r}^2 + 2\m \vec{y}\cdot \partial_\tau \vec{x}],
\ee
which is exactly solvable. 

Finally we consider motion along the $\p_1$ coordinate, as 
before we introduce
\be
\tix^+= {1\over 2}(t+{1\over \sqrt{6}}\p_1), \quad  
\tix^-= {1\over 2}(t-{1\over \sqrt{6}}\p_1)
\ee
and further taking the limit
\begin{eqnarray}
x^+&=&\tix^+, \quad x^-=R^2 \tix^-, \quad \r={r\over R},  \nonumber \\
\te_i &=& {\pi\over 2}+\sqrt{6} { y_i\over R},\quad  
\p_2 = \sqrt{6}{x_2\over R},
\quad  \psi =3 {z\over R}.
\end{eqnarray}
we obtain the following metric resembling the motion of a particle in a magnetic 
field. 
\begin{eqnarray}
ds^2&=&-4dx^+ (dx^- +\m y_1 dz)-\m^2 (\vec{r}\,{}^2+2y_1^2)(dx^+)^2
+ d\vec{r}\,{}^2+ d\vec{y}{}^2 + dx_2^2 +dz^2, \nonumber \\
F_5&=&\F_5+ *\F_5, \quad \F_5\sim \m \,dx^+\wedge dz\wedge dy^1\wedge dy^2 
\wedge dx^2,
\end{eqnarray}
where $\vec{y}=(y_1,y_2)$.

\subsection{PP wave limit on $AdS_5 \times T^{p,q}$}

In this subsection we discuss the analogous problem for the motion of a particle in 
$T^{p,q}$ space. We thus consider a solution to the IIB equations 
of motion given by 
$AdS_5\times T^{p,q}$ where the metric on $T^{p,q}$ is given by
\begin{eqnarray}
ds^2_{T^{p,q}}&=&\l_1^2(d\te_1^2+\sin^2 \te_1d\p_1^2)
+\l_2^2(d\te_2^2+\sin^2 \te_2d\p_2^2) \nonumber \\
&+&\l^2(d\psi+p\cos \te_1 d\p_1 +q\cos\te_2 d\p_2)^2.
\end{eqnarray}
This space is Einstein provided:
\be
{1\over \l_1^2}\big[1-{\l^2p^2\over 2\l_1^2}\big]=
{1\over \l_2^2}\big[1-{\l^2q^2\over 2\l_2^2}\big]=
{\l^2\over 2}\big[{p^2\over \l_1^4}+{q^2\over \l_2^4}\big]
\ee
In this sense the metric considered in the previous section is a particular case
of this more general space corresponding to $p=q=1, \l_1^2=\l_2^2=1/6 $
and $\l^2=1/9$. The element of this class of spaces that have attracted 
the most attention is precisely $T^{1,1}$ since it is the only space that
provides a supersymmetric solution of IIB \cite{romans}. 

Next we perform the following coordinate transformation 
\begin{eqnarray}
\tix^+&=& {1\over 2}\bigg[t+\L\l(\psi + \p_1 + \p_2)\bigg], \nonumber \\
\tix^-&=& {1\over 2}\bigg[t-\L\l{1\over 3}(\psi + \p_1 + \p_2)\bigg], \nonumber \\
\Phi_1&=&\p_1- {\l p\over 4\L\l_1^2}\bigg[t+\L\l(\psi + \p_1 + \p_2)\bigg], \nonumber \\
\Phi_2&=&\p_2- {\l q\over 4\L\l_2^2}\bigg[t+\L\l(\psi + \p_1 + \p_2)\bigg].
\end{eqnarray}
Here we have introduced a constant $\L$ which is used to match the radii of
$AdS_5$  and $T^{p,q}$. Taking the large $R$ limit in the form 
\be
x^+=\tix^+, \quad x^-=R^2 \tix^-, \quad \r={r\over R}, 
\quad \te_i = {r_i\over \L \l_i R}.
\ee
 we obtain
\begin{eqnarray}
ds^2&=&-4dx^+ dx^- -  \m^2 (r^2 + {\l^2p^2\over 4\L^2\l_1^4}r_1^2 
+{\l^2q^2\over 4\L^2\l_2^4} r_2^2)( dx^+)^2 \nonumber \\
&+& dr^2+r^2 d\O_3^2+ dr_1^2
+r_1^2d\p_1^2 +dr_2^2 +r_2^2 d\p_2^2.
\end{eqnarray}
Note that each  pair $(r_i,\p_i)$ parametrizes an  ${\bf R}^2$. At the 
string theory level we can identify, in principle, three different masses
for the eight transverse coordinates. Although here we are considering 
a conformal background, this splitting is characteristic 
in  nonconformal situations, where different coordinates get different 
masses due to the introduction of nonconformality as will be shown in the 
next section.

\section{Nonconformal Cases}
One of the most striking features of the AdS/CFT is that it allows to go
beyond its original statement about the duality between IIB string
theory in $AdS_5\times S^5$ and ${\cal N}=4$ large $N$ super
Yang-Mills. From phenomenological perspective  backgrounds that are not
conformal invariant play a very important role. In this section we
consider the Penrose limit of  three different types of  nonconformal 
backgrounds. Namely, we consider the Schwarzschild black hole in 
$AdS_5$ complemented by $S^5$ and a nontrivial 5-form; in this case
the scale in the problem is provided by the mass of the Schwarzschild 
black hole. We also consider the Maldacena limit of D3-branes placed
at the origin of the
small resolution of the conifold, which is 
a natural deformation of the $AdS_5\times T^{1,1}$ solution considered 
before. Finally in this section we consider the simplest of the 
theories with varying 5-form flux - 
the Klebanov-Tseytlin solution \cite{kt}. 
As oppose to the conformal cases discussed before non of this deformation 
allows for exact solvability at the string theory level due to an
explicit  dependence on the metric on the $x^+$ coordinate. However, it
is possible to extract some information about the field theory side.

\subsection{Penrose Limit of the Schwarzschild Black Hole in AdS}
In this section we consider the Penrose limit for the Schwarzschild
black hole in AdS. This deformation of the AdS/CFT, when analytically continued
to Euclidean time is related to a  nonzero temperature deformation 
in the field theory
side. However, here we consider strictly the Lorentzian signature since it
is a requirement for the implementation of the Penrose limit \cite{penrose}. 
The metric we consider in this subsection is 

\be
R^{-2}ds^2=-r^2(1-{r_0^4\over r^4})dt^2 + {dr^2\over r^2(1-{r_0^4\over r^4})}
+ r^3 d\O_3^2 + d\psi^2 + \sin^2\psi d\O_4^2.
\ee
Before applying to Penrose limit we bring the metric into a convenient
form following \cite{penrose},\cite{tratado}. We study the null geodesic
determined by the following effective Lagrangian
\be
{\cal L}= -r^2(1-{r_0^4\over r^4})\dot{t}^2 +\dot{\psi}^2+{\dot{r}^2\over
r^2(1-{r_0^4\over r^4})},
\ee
where dot represents derivative respect to the affine parameter. Since
the effective Lagrangian does not depend explicitly on the coordinates
$t$ and $\psi$ we have two integrals of motion:
\be
\dot{\psi}=\m, \quad \dot{t}={E\over r^2(1-{r_0^4\over r^4})}.
\ee
For null geodesics ${\cal L}=0$ we obtain an equation for $r$
\be
\dot{r}^2+\m^2r^2(1-{r_0^4\over r^4})=E^2.
\ee
We choose the affine parameter $u$ as part of a new coordinate system
$(u,v,\phi)$ in which we can enforce $g_{uu}=g_{u\p}=0$ and
$g_{uv}=1$. In these coordinates the metric takes the form
\begin{eqnarray}
R^{-2}ds^2&=&2dudv+2\m^2r^2(1-{r_0^4\over r^4})dv d\p -r^2(1-{r_0^4\over
r^4})dv^2 +\big[1-(1-{r_0^4\over r^4})\big]d\p^2 \nonumber \\
&+&r^2 d\O_3^2 +\sin^2(\p + \m u) d\O_4^2,
\end{eqnarray}
with
\be
{dr\over du}=\sqrt{1-\m^2r^2(1-{r_0^4\over r^4})}.
\ee

We now perform the Penrose limit by sending $R\to \infty$ with 
\be
u\to u, \quad v\to {v\over R^2}, \quad Y^i \to {y^i\over R},
\ee
where $Y^i$ represent a subset of the rest of the coordinates. We find
the Penrose limit for the Schwarzschild black hole in AdS to be 
\be
ds^2 =2dudv +(1-\m^2 \r^2(u))d\p^2 + r^2(u) ds^2({\bf R}^3) + \sin^2 (\m u)
ds^2({\bf R}^4),
\ee 
where 
\be
\r^2=r^2(1-{r_0^4\over r^4}).
\ee
As noted in \cite{tratado} for $(r_0=0)$ the Penrose limit is flat space
if $\m \equiv 0$; for $r_0=0$ we recover the result corresponding to the
Penrose limit in $AdS_5 \times S^5$. The above result is presented in  Rosen
coordinates we can transform the metric into the usual Brinkman
coordinates in which case we obtain:
\be
ds^2=-4dx^+dx^- -\m^2 \bigg[(1-{3r_0^4\over r^4})\p^2 +(1+{r_0^4\over
r^4})z_3^3+z_4^2\bigg](dx^+)^2 + d\p^2 + dz_3^2 + dz_4^2,
\ee
where $z_3$ parametrizes ${\bf R}^3$ and $z_4$ parametrizes ${\bf
R}^4$. In this coordinate system
\be
x^+={1\over
2\m}\arctan\bigg[{2\m^2r^2-1\over 2\m\sqrt{r^2-\m^2r^4+\m^2r_0^4}}\bigg].
\ee 
Note that the coordinates parametrizing ${\bf R}^4$ originate from 
the $S^5$ and are not affected by the nonconformality introduced by the
Schwarzschild-AdS black hole. The masses of the bosons that have been 
affected are to those directions lying within AdS. 

\subsection{The small resolution of the conifold}
In this subsection we consider a very different type of nonconformality
from the previous section. Namely, we consider the Maldacena limit of
regular D3 branes on the small resolution of the conifold. A natural
scale is introduced in the problem by the radius (minimal volume) of the
nonvanishing $S^2$. The field theory interpretation of this solution was
discussed by Klebanov and Witten in \cite{kw2}, the supergravity
construction was presented in \cite{pt}. The near horizon limit is of 
the standard D3-brane form
\be
ds^2=h^{-1/2}\eta_{\m\n}dx^\m dx^\n + h^{1/2} ds_6^2,
\ee
where $ds_6^2$ is  the metric of the small resolution of the conifold
\begin{eqnarray}
ds_6^2&=&{ \k^{-1}}dr^2+ \de  \k\,\,  r^2 e_{\psi}^2 +
\si r^2\left(e_{\te_1}^2+e_{\p_1}^2\right)+\si
 ({r^2} +
6a^2)\left(e_{\te_2}^2+e_{\p_2}^2\right)\ , \nonumber \\
\k &\equiv &{r^2+9a^2\over r^2+6a^2}, \quad e_{\te_i}=d\te_i, 
\quad e_{\p_i}=\sin\te_i d\p_i \ .
\end{eqnarray}
As usual the warp factor is 
 a  harmonic function  on the transverse 6-d space:
\begin{equation}
h={R^4\over 9a^2 r^2}\bigg[1-{r^2\over 9 a^2}\ln(1+{9a^2\over
r^2})\bigg].
\end{equation}
Proceeding as in the previous cases, we  study the 
geodesic line along $(r,t,\psi)$ in order to find a natural 
transformation into coordinates $(u,v,x)$ convenient to perform the 
Penrose limit. Namely, we consider the null geodesic described by

\be
{\cal L}=-{3ar\over \sqrt{1-{r^2\over 9 a^2}\ln(1+{9a^2\over
r^2})}}\dot{t}^2 + {\sqrt{1-{r^2\over 9 a^2}\ln(1+{9a^2\over
r^2})}\over 3ar}\k^{-1} \dot{r}^2 + {\sqrt{1-{r^2\over 9 a^2}\ln(1+{9a^2\over
r^2})}\over 3a}\k \,r\, \dot{\psi}^2.
\ee

The equations following from this Lagrangian are:
\begin{eqnarray}
\dot{t}&=& {\sqrt{1-{r^2\over 9 a^2}\ln(1+{9a^2\over
r^2})}\over 3ar}, \nonumber \\
\dot{\psi}&=&{3a\m\over \k\,r \,\sqrt{1
-{r^2\over 9 a^2}\ln(1+{9a^2\over r^2})}}, \nonumber \\
\dot{r}^2&=&{r^2+9a^2\over r^2+6a^2}- {9a^2 \m^2 \over 1
-{r^2\over 9 a^2}\ln(1+{9a^2\over r^2})}.
\end{eqnarray}
We take the Penrose-G\"uven limit as
\be
u=u , \quad v\to {v\over R^2}, \quad x\to {x\over R}, 
\quad \te_i = \sqrt{6} {r_i\over R},
\ee
resulting in 
\begin{eqnarray}
ds^2&=&2dudv + {\k\, \L\over 3a}
\bigg[1-{9a^2\m^2\over \k\,\L^2}\bigg]dx^2 
+{\L r \over 3a} ds^2({\bf R}_1^2)
+{\L r \over 3a}(1+{6a^2\over r^2}) ds^2({\bf R}_2^2), \nonumber \\
\L&=&\sqrt{1-{r^2\over 9a^2}\ln (1+{9a^2\over r^2})},
\end{eqnarray}
where ${\bf R}_i^2$ represents the ${\bf R}^2$ parametrize 
by $(r_i, \p_i)$. This metric can be brought to Brinkman coordinates
following the prescription of \cite{tratado}. 
As in the previous case this metric smoothly goes 
to the maximally supersymmetric pp-wave as the conformality parameter 
goes to zero $(a\to 0)$.

\subsection{The Klebanov-Tseytlin solution}
The Klebanov-Tseytlin solution describes the geometry of a collection of
regular and fractional branes on the conifold \cite{kt}. It contains 
a naked
singularity in the IR where it must be replaced by the
Klebanov-Strassler solution corresponding to the replacement of the 
conifold by the  deformed 
conifold \cite{ks}. Although not completely accurate, the KT 
geometry provides a simple description of the supergravity dual of the 
breaking of conformal
invariance in the field theory by the introduction of fractional
branes, it is also computationally a lot more manageable than the 
corresponding background for the 
deformed conifold metric. We thus proceed, with caution, to study the
Penrose limit in the KT solution. 

In this section we will use the Poincare coordinates of $AdS$ that are
naturally related to the standard D3-brane solution. This approach in
principle obscures the relation of the time coordinate to the global
time coordinate that we used in the previous sections and that was used
in \cite{bmn}. However, as was explicitly shown in \cite{tratado}, the
end result is the same limit and this completely shows the equivalence
of both routes. The metric of the KT solution is

\begin{eqnarray}
ds^2&=&h^{-1/2}\eta_{\m\n}dx^\m dx^\n + h^{1/2}\big[dr^2 +r^2ds_{T^{1,1}}^2\big],
\nonumber \\
h&=&{R^4\over r^4}(1+P\ln({r\over r_0})).
\end{eqnarray}
For the precise normalizations we refer the reader to
\cite{kreview}. For us it will be important the $P$ is proportional to
the number of fractional D3-branes and that it is natural to consider
it small with respect to the number of regular D3-branes which is
proportional to $R^4$. After a rescaling of the radial coordinate, and
similarly $r_0$, we bring the metric to the form
\be
R^{-2}ds^2={r^2\eta_{\m\n}dx^\m dx^\n\over \sqrt{1+P\ln ({r\over
r_0})}}+ r^{-2}\sqrt{1+P\ln ({r\over r_0})}\,\,dr^2 + 
\sqrt{1+P\ln ({r\over r_0})}\,\,ds_{T^{1,1}}^2.
\ee
As in the case of $AdS_5 \times T^{1,1}$ we concentrate on motion along
a geodesic given by $\psi + \p_1 +\p_2$ in a small neighborhood of
$\te_i=0$. The effective Lagrangian from which the geodesic equation
follows is 
\be
{\cal L}=-{r^2 \dot{t}^2\over \sqrt{1+P\ln ({r\over r_0})}} +
r^{-2} \sqrt{1+P\ln ({r\over r_0})}\,\,\dot{r}^2 + 
\sqrt{1+P\ln ({r\over r_0})} \,\,\dot{\psi}^2,
\ee
where dot represents derivative with respect to the affine parameter u. 
Since the Lagrangian does not  explicitly depend on $t$ and $\psi$ we have
two integral of motion:

\be
\dot{t}={E\over r^2} \sqrt{1+P\ln ({r\over r_0})}, \quad 
\dot{\psi}= \m/\sqrt{1+P\ln ({r\over r_0})}.
\ee
From these two relations we find that 
\be
\dot{r}^2 +{\m^2r^2\over 1+ P\ln({r\over r_0})}=E^2.
\ee
Our aim is, following 
\cite{penrose,gueven,tratado}, to find  new 
coordinates $(u,v,x)$ satisfying $g_{uu}=0, \,\,
g_{uv}=1$ and $ g_{u x}=0$, in which the Penrose-G\"uven limit is
naturally taken. A simple solution satisfying this transformation was
given in \cite{tratado} and can be naturally extended to the case under
consideration 
\be
\partial_u=\dot{r}\partial_r + \dot{t}\partial_t+\dot{\psi}
\partial_{\psi}, \quad \partial_v=-{1\over E} \partial_t, \quad \partial_x =
\m\partial_t + E\partial_\psi.
\ee
From now on we set $E=1$, as in the previous cases. 
This system can be integrated. After taking
the Penrose limit following 
\be
u \to u , \quad v \to {v\over R^2}, \quad \te_i = \sqrt{6}{r_i\over R}, \quad x\to
{x\over R},
\ee
we obtain the following metric 
\begin{eqnarray}
ds^2&=&2 dudv + {r^2\over \sqrt{1+P\ln ({r\over r_0})}} dx_3^2 +
\sqrt{1+P\ln ({r\over r_0})}\bigg[1-{\m^2r^2\over 1+P\ln({r\over
r_0})}\bigg]dx^2 \nonumber \\
&+& \sqrt{1+P\ln ({r\over r_0})}\big[dr_1^2
+r_1^2d\p_1^2 +dr_2^2 +r_2^2 d\p_2^2 \big],
\end{eqnarray}
where $r$ and $u$ are related according to 
\be
u=\int{dr\over \sqrt{1-\m^2r^2/(1+P\ln({r\over r_0}))}}.
\ee
Note that in the particular case of $P=0$ we find that the above expression
can be explicitly integrated $(r=\m^{-1}\sin \m  u)$ giving the pp-wave 
in Rosen coordinates
\cite{tratado}. One of the most interesting characteristics of the
KT-type backgrounds is the dependence on the 5-form on the radius,
associated with the RG-cascade. In the present context this feature
remains 
\be
F_5=\F_5 + * \F_5, \quad \F_5 \approx (1+P\ln({r\over r_0}))\dot{\psi}du
\wedge dr_1\wedge r_1 d\p_1 \wedge dr_2 \wedge r_2d\p_2.
\ee
Similarly some of the components of the 3-form fields survive the
Penrose-G\"uven limit
\begin{eqnarray}
B_2&\sim &P \ln({r\over r_0})( dr_1\wedge r_1d\p_1-dr_2\wedge
r_2d\p_2), \nonumber \\
F_3&\sim& P \dot{\psi} du \wedge (dr_1\wedge r_1d\p_1-dr_2\wedge r_2 d\p_2)
\end{eqnarray}
\subsection{Comments on the general structure of the Penrose 
limit in nonconformal backgrounds}
We have seen that in all the three examples that we discussed the 
nonconformality parameters appear naturally as a perturbation of the 
metric away from the exactly solvable pp-wave limit discussed in 
\cite{metsaev,metse} 
\be
ds^2=-4dx^+dx^- + d\vec{z}{}^2 
-\m^2 \bigg[\sum\limits_i(1+\epsilon f_i(x^+))z_i^2\bigg](dx^+)^2,
\ee
where $\epsilon$ is the nonconformality parameter. The collection of 
functions $f_i(x^+)$ characterize the form of the ``mass'' 
deformation  for the coordinate $z_i$; in the case of the 
Schwarzschild black hole we note that $f_i=0$ for the directions
within the $S^5$, as expected.  From the string theory point of 
view we have that  since the equation of motion for 
$x^-$ implies that $\Box x^+=0$ we can fix the world sheet diffeomorphism 
invariance by choosing the light cone $x^+=p^+ \tau$. The resulting gauge 
fixed string  theory action will then be interpreted as a theory of eight 
massive fields with time-dependent mass. 

An interesting observation is that the fields $z_i$ can not appear to 
order higher than two, that is to say, the action is always quadratic in 
the field $z_i$. This can be seen by recalling that in the 
Penrose-G\"uven limit one rescales $z_i \to z_i/R$ and then multiplies 
the metric by $R^2$. 

The near horizon limit of nonconformal  Dp-branes has been  discussed
in \cite{cobi}, their Penrose-G\"uven limit has been  presented in
\cite{tratado}. It is interesting to note that they also fall into the
general form discussed here, in the sense that they  have and
$x^+$-dependent mass function.

\section{${\cal N}=1$ Super Conformal Field Theory interpretation}
\label{field}

In this section we attempt to find a field theory  interpretation to
the limits taken in section 1 for the conifold geometry
 of
$AdS_5\times T^{1,1}$. 
 We follow closely  the procedure used  in
\cite{bmn}, in the Penrose  limit of section (1.1), namely, we
first  identify  the field theory operators that correspond to the
ground  state with large
 $p^+$ and to the first excited level. We
then make some preliminary observations about  
 the possibility to
have a perturbative expansion that will reproduce the string
Hamiltonian. We end by discussing the operator identifications of the
magnetic 
 cases of section (1.2).  It is crucial for us to relate 
isometries of the $T^{1,1}$ space with charges of the operators of
the CFT. To make that connection completely transparent we start
with  reviewing  the  construction of the conifold metric following
Candelas and De la Ossa  \cite{candelas} and some facts  about the
superconformal field theory  discussed by Klebanov and Witten
\cite{kw} (we will also rely on the analysis of \cite{ceresole}).  
 
\subsection{Review of the the conifold and the superconfomal theory of D3
branes at the conifold singularity}

The conifold is defined  by the following quadric in ${\bf
C}^4$:
\be
\label{defining}
\sum_{i=1}^4 w_i^2=0.
\ee
This equation can be written as
\begin{equation}\mbox{det}\ \W=0\ , \ \ \ \  
 {\rm i.e.} \ \ \ \ Z_1 Z_2-Z_3Z_4=0\
,
\ee \be
\label{conifold}
\W=
{1\over \sqrt{2}}\left(
\begin{array}{cc}
w_3+iw_4&w_1-iw_2\\
w_1+iw_2&-w_3+iw_4
\end{array}
\right)
\equiv
\left(
\begin{array}{cc}
Z_1&Z_3\\
Z_4&Z_2
\end{array}
\right).
\end{equation}
Equation (\ref{defining}) has an $SO(4)$ symmetry that is usually
treated as an $SU(2)\times SU(2)$. There is also a $U(1)$ symmetry given
by
\be
w_i\to e^{i\a} w_i.
\ee
This last symmetry was identified with the $U(1)_R$ in the gauge theory
side based on the fact that the holomorphic 3-form can be written as
\be
\O={dw_1\wedge dw_2\wedge dw_3 \over w_4},
\ee
and therefore has charge two under this $U(1)$ symmetry. Moreover, since
the chiral superspace coordinate transform as $\O^{1/2}$ we obtain that
it naturally has charge one. 

To write an explicit metric on the conifold we need to find a general
solution to equation (\ref{conifold}) and assume that the K\"ahler
potential depends only on the radial coordinate which is defined as: 
$r^2=\mbox{tr} (\W^{\dagger} \W)$. The most general solution can be
found by acting on a particular solution with elements of $SU(2)\times
SU(2)$. Namely, given a particular solution $Z_0$ we construct $\W$ as  
\be
{\W\over r}=L Z_0 R^{\dagger}=
\left(
\begin{array}{cc}
a&-\bar{b}\\
b&\bar{a}
\end{array}
\right)
\left(
\begin{array}{cc}
0&1\\
0&0
\end{array}
\right)
\left(
\begin{array}{cc}
\bar{k}&\bar{l}\\
-l&k
\end{array}
\right)
\ee
where $|a|^2 +|b|^2=|a|^2 +|b|^2=1$ and they can be parametrize as
$a=\cos{\te_1\over 2}\exp {i\over 2}(\psi_1+\p_1),\,\, 
b=\sin{\te_1\over 2}\exp {i\over 2}(\psi_1-\p_1)$ and similarly for $k$
and $l$. With this choice of parametrization of $SU(2)$ we find
\begin{eqnarray}
Z_1&=&-r \cos{\te_1\over 2}\sin{\te_2\over 2}\exp {i\over 2}(\psi +
\p_1-\p_2), \nonumber \\
Z_2&=&r \sin{\te_1\over 2}\cos{\te_2\over 2}\exp{i\over 2}(\psi-
\p_1+\p_2), \nonumber \\
Z_3&=&r \cos{\te_1\over 2}\cos{\te_2\over 2}\exp{i\over 2}(\psi +
\p_1+\p_2), \nonumber \\
Z_4&=&-r \sin{\te_1\over 2}\sin{\te_2\over 2}\exp{i\over 2}(\psi -
\p_1-\p_2), 
\end{eqnarray}
where  $\psi=\psi_1+\psi_2$. Looking at equation (\ref{conifold}) we see
that the $Z's$ are linear combinations of the $w's$ and thus in order for
the latter to have $R$-charge one we must identify that symmetry with shifts of 
$\psi=2\beta$, precisely as in  \cite{kow}.

\subsection{Identification of lowest `` string modes'' of the KW model } 
The lowest components of the  
chiral superfields of the KW model  are related to  the conifold parameters in the following way 
 \cite{kw} 
\be  
Z_1=A_+B_+, \quad Z_2=A_-B_-,\quad Z_3=A_+B_- \quad Z_4=A_-B_+.  
\ee  
where $(A_-,A_+)$ and $(B_-,B_+)$ are doublets of $SU(2)_A$ and $SU(2)_B$ global symmetries  
respectively and carry a $U(1)_R$ charge of $1/2$. 
The question now is how to characterize these fields in the sector that corresponds to the Penrose limit. 
In \cite{bmn} the light-cone Hamiltonian was taken to be 
$2P^-=\Delta-J$ where $J$ is the generator of 
rotations in the $\psi$ direction, $J=-i\partial_\psi$. 
Following (\ref{t11}) and the discussion above about the identification of $\psi=2\beta$, 
 it is natural to identify $J$ for the  $T^{1,1}$ case as follows 
\be 
J=-i[1/2\partial_\psi+\partial_{\phi_1} +\partial_{\phi_2}]. 
\ee  
In the following table we  classify the fields of the KW theory with  respect 
to 
their gauge transformations, their $U(1)_A\times U(1)_B\times U(1)_R$ 
charges, where 
$U(1)_A$ and $U(1)_B$ associate with the $ 
T_3$ generators of  $SU(2)_A$ and $SU(2)_B$ respectively, 
their $J= 1/2 U(1)_R+U(1)_A +U(1)_B$ charges and their conformal dimension. 
In addition the table contain composite operators that carry $\Delta- J =0,1$.

\begin{center}  
\begin{tabular}[h]{|l||c|c|c|c|c|c|c|c|}  
\hline  
                   & $SU_L(N)$  &$SU_R(N)$ & $U_A(1)$ & $U_B(1)$&$U_R(1)$  & $J $& $\Delta$&$\Delta-J$ \\  
\hline   
\hline   
$(A_+,\psi^A_+)$ & $N$ & $\bar N$ & $1/2$ & 0 & $(1/2,-1/2)$ & $(3/4,1/4)$ &$(3/4,5/4)$& $ (0,1)$ \\  
\hline   
$(A_-\psi^A_-)$ & $N$ & $\bar N$ & $-1/2$ & 0 & $(1/2,-1/2)$ & $(-1/4,-3/4)$ &$(3/4,5/4)$& $ (1,2)$ \\  
\hline   
$(B_+\psi^B_+)$ & $\bar N$ & $ N$ & $0$ & $1/2$ & $(1/2,-1/2)$ & $(3/4,1/4)$ &$(3/4,5/4)$& $ (0,1)$ \\  
\hline   
$(B_-\psi^B_-)$ & $\bar N$ & $ N$ & $0$ & $-1/2$ & $(1/2,-1/2)$ & $(-1/4-3/4)$ &$(3/4,5/4)$& $ (1,2)$ \\  
\hline  
\hline   
$\lambda_L$ & $adj $ & $ 1$ & $0$ & $0$ & $1$ & $1/2$ &$3/2$& $ 1$ \\  
\hline   
$\lambda_R $& $1 $ & $ adj$ & $0$ & $0$ & $1$ & $1/2$ &$3/2$& $ 1$ \\  
\hline\hline  
$Z\equiv A_+B_+$ & $adj\oplus 1$ & $adj\oplus 1$ & $1/2$ & 1/2 & $1$ & $3/2$ &$3/2$& $ 0$ \\  
\hline  
\hline   
$\phi_1\equiv A_+B_-$ & $adj \oplus 1$ & $adj \oplus 1$ & $1/2$ & -1/2 & $1$ & $1/2$ &$3/2$& $ 1$ \\  
\hline   
$\phi_2\equiv A_-B_+$ & $adj \oplus 1$ & $adj \oplus 1$ & $-1/2$ & +1/2 & $1$ & $1/2$ &$3/2$& $ 1$ \\  
\hline\hline   
  
$\psi_1\equiv A_+\psi^B_+$ & $adj \oplus 1$ & $adj \oplus 1$ & $1/2$ & -1/2 & $0$ & $1$ &$2$& $ 1$ \\  
\hline   
$\psi_2\equiv B_+\psi^A_+$ & $adj \oplus 1$ & $adj \oplus 1$ & $-1/2$ & +1/2 & $0$ & $1$ &$2$& $ 1$ \\  
\hline  
\end{tabular}  
\end{center}  
  
From the table it is clear that in a similar manner to the ${\cal N}=4$ 
case \cite{bmn}, 
the natural candidate in the KW model for  the operator 
that corresponds to the 
light-cone ground state is 
\be
{1\over \sqrt{J}N^{J/2}}Tr[(A_+B_+)^J] 
\ee
The bosonic operators associated with $\Delta -J=1$ are the $D_i Z$ and $\phi_1$ and $\phi_2$ 
defined in the table. The missing 2 bosonic operators are 
associated with non chiral operators composed from complex 
conjugates of the basic bosonic fields as was pointed 
out in \cite{gubmit} 
\footnote{ We thank O. Aharony for pointing this out to us.} 
As for the fermionic operators at ``level''one  we 
have the operators defined in the table as $\psi_1$  
and $\psi_2$ and in addition there are the gauginos 
and the missing operators aren again associated 
with the corresponding  non chiral operators.

\subsection{On the ``strings'' of the KW model} 
Since the Penrose limit of the   $AdS_5 \times T^{1,1}$ is identical to  that of the   
 $AdS_5 \times S^5$, it is clear that the bosonic part of the  
light-cone string action  associated  with the  
former limit  takes the form \cite{bmn}  
 \be\label{stringaction} 
S= {1\over 2\pi\alpha'}\int d^2 \sigma [{1\over 2}\partial_\alpha \vec{z} \partial^\alpha \vec{z} 
-\mu^2\vec{z}^2]  
\ee 

The question now is whether  one can  show  that  the  KW    
field theory in the Penrose limit  admits a string bit picture 
\cite{thorn} which is   
governed by a discretized Hamiltonian that flows in the continuum limit to the Hamiltonian 
associated with  (\ref{stringaction}).  
The Penrose limit now associates with 
\be
g_s^2N\rightarrow \infty \qquad 
{g_s^2N\over J^2}\ fixed  \qquad \Delta -J\   fixed,
\ee
Note that it is $g_s$ and not $g_{YM}^2$ which is involved in the limit since the later becomes large
at the IR fixed point. $g_s$ maps in the field theory language, in a manner that is not completely 
understood to $\lambda$ the coupling of the superpotential which is given by  
\be
W={\lambda\over 2}\ep^{ij}\ep^{kl}Tr[A_iB_kA_jB_l]  
\ee
with $i=+,-$. 
The corresponding scalar potential takes the form 
\be
V=G^{r\bar s}\partial_r W\partial_{\bar s}\bar W = Tr[G^{A^\dagger_+A_+} (B_+A_-B_-)(B^\dagger_+A^\dagger_-B^\dagger_-)]+...
\ee
where the ... corresponds the other terms  in $G^{r\bar s}$.
 The k\"ahler metric  \cite{kw}
can be rewritten in terms of the $A_i$ and $B_i$ fields, inverted 
and inserted in the expression  
for the potential (which has total dimension four). 

A key issue is obviously whether one can find a perturbative 
expansion of this  interaction potential. 
It is hard to believe that such a perturbation is possible 
especially since the theory is at strong
gauge coupling. On the other hand as stayed above the final 
answer should converge to a free light-cone ``massive'' string.  
 The resolution of the puzzle how such a complicated potential can lead to  
 the same continuum result as for the  $AdS_5 \times S^5$ may  
involve some non-trivial 
 map that  
will transform the  interaction potential into  
$Tr[\hat Z \hat \phi \bar {\hat Z}\bar{\hat \phi}]$ 
where $\hat Z$ and  $ \hat \phi$ are dimension one 
operators of $\Delta -J$ equal to $0$ and $1$ respectively, like for 
instance $\hat Z=Z^{2/3}, \ \hat \phi= \phi/Z^{1/3}$. 
This type of  construction is under current investigation.   
 
\subsection{ String modes in the magnetic limit}
\label{magnetic}
In spite of the fact that the Penrose limits of (\ref{n4}) and
(\ref{mag}) are related  by coordinate transformation we will argue 
that the corresponding light-cone Hamiltonians are different and 
therefore determine different projections in the space of 
operators.  In analogy to using 
$H_{lc}=2P^-=\Delta-\big[(1/2)U(1)_R+U(1)_A+U(1)_B\big]$ 
for the Penrose limit of subsection  2.1 
we identify $J=\alpha U_R(1)$, where $\alpha$ is some numerical 
constant 
 for  the limit where $\tix^+= {1\over 2}(t+{1\over 3}\psi)$. It is 
thus clear that all the $A_i$ and $B_i$  will have the same 
eigenvalue of $2P^-$. In particular for $\alpha=3/2$ we have that 
$\Delta-J= 0$   for all the scalar fields and the gauginos.  In 
this case the vacuum state will be highly degenerate since there 
will be many operators that are analogous to $Tr[Z^J]$.
This high degeneracy may  be related to the degeneracy of the 
corresponding Landau levels. 
The fermions $\ps^A_\pm$ and $\ps^B_\pm$ have  $\Delta-J= 2$

For the case that $\phi_i$ is replacing $\psi$ in the definition 
of $x^+$ then naturally $J=\alpha U(1)_A$. In this case there is 
less degeneracy since $J$  distinguishes  between $A_+$ and $A_-$ but 
since the $B_i$ carry  zero $J$ charge there is still some degeneracy. 
For instance for $\alpha=3$ both
$Tr[A_+B_+]$ and $Tr[A_+B_-]$ correspond to the string ground state. 

A similar situation is also encountered in $AdS_5\times S^5$. Using 
different linear combinations of the $U(1)$'s, implies using different $J$'s 
and therefore different light-cone Hamiltonians $H_{lc}$. In particular, it is 
possible to get the magnetic case by  taking the sum of the three $U(1)$'s inside
$SU(4)$.

\begin{center}
{\large  Acknowledgments}
\end{center}
We would like to thank D. Berenstein, J. Maldacena,
S. Mukhi  and C. Thorn for conversations; we also thank J. Gomis, 
G. Papadopoulos and A. Tseytlin for comments on a previous 
version of the manuscript. We are especially grateful 
to O. Aharony for very useful discussions. The research of 
JS was supported in part 
by the US-Israel Binational Science
Foundation, by GIF -- the German-Israeli Foundation for Scientific 
Research, and by the Israel Science Foundation.

\appendix{Local versus global existence of geodesic congruence}
In this appendix we explain the intrinsically local character of some of
the limits taken in
the main body of the test. Consider the construction of null geodesic in
the plane given by $(t,\r,\psi)$ in any of the geometries discussed
previously of for that matter even in $AdS_5\times S^5$ we concentrate
on the part of the metric having the following form:
\be
ds^2 = -dt^2 \cosh^2\r + d\r^2 + d\psi^2.
\ee
The equation of the geodesic in this coordinates follows from the
Lagrangian
\be
{\cal L}= -\dot{t} ^2 \cosh^2\r^2 +  \dot{\r}^2 + \dot{\psi}^2,
\ee
where the dot means derivatives with respect to the affine parameter
$u$. Since there is not explicit dependence on $\psi$ or $t$ we have
two  integral of motion:
\be
\dot{\psi}=\m, \quad \dot{t}=E/\cosh^2\r.
\ee
substituting in the condition of null geodesic ${\cal L}=0$ we get 
\be
\dot{\r}^2 + \m^2 = {E^2\over \cosh^2\r}.
\ee
This equation shows the local character of such geodesic. Namely, in a
neighborhood of $\r=0$ there is always a solution to this
equation with nonzero $\m$ and therefore including $\psi$. However if we
allow  $\r$ to be very large we find that the above equation can always
be falsified unless we take $\m \equiv 0$ in which case the solution is 
\be
\m \equiv 0,\quad  \sinh \r =E u, \quad t = \arctan Eu - v/E.
\ee
which shows that away from a neighborhood of $\r=0$ the geodesic line is
completely independent of $\psi$ and lies wholly within
$AdS_5$. Further, is was shown in \cite{tratado}, explicitly and using
the hereditary properties of Penrose limits, that  the Penrose limit on
$AdS$ always results in flat space.
The key point exploited in the body of the paper is that we can set the
size of the small neighborhood of $\r=0$ by rescaling by $R$ and up  to
fourth order in $1/R$ there is a null geodesic nontrivially including a
dependence on $\psi$, which is precisely the one used in the body of the
paper.

\end{document}